\documentclass[11pt, a4paper]{article}
\usepackage[left=4cm, right=4cm, 
]{geometry}

\usepackage{local_config}

\usepackage{titling}
\usepackage{abstract}

\usepackage{authblk}

\title{Non-random network connectivity comes in pairs\vspace{-2ex}}
\date{}
\author[1,2,*]{Felix Z.~Hoffmann}
\author[1]{Jochen Triesch}

\affil[1]{Frankfurt Institute for Advanced Studies (FIAS), Johann Wolfgang Goethe University, Frankfurt am Main, Germany}
\affil[2]{International Max Planck Research School for Neural Circuits, Max Planck Institute for Brain Research, Frankfurt am Main, Germany\vspace{3ex}}
\affil[*]{Email: hoffmann@fias.uni-frankfurt.de\vspace{-13.5ex}}

\begin{document}



\maketitle
\begin{abstract}
Overrepresentation of bidirectional connections in local cortical networks has been repeatedly reported and is in the focus of the ongoing discussion of non-random connectivity. Here we show in a brief mathematical analysis that in a network in which connection probabilities are symmetric in pairs, $P_{ij} = P_{ji}$, the occurrence of bidirectional connections and non-random structures are inherently linked; an overabundance of reciprocally connected pairs emerges necessarily when the network structure deviates from a random network in any form.
\end{abstract}
\vspace{0.2cm}

\section*{Introduction}



Increasing evidence shows that cortical microcircuitry is highly structured \cite{Song2005,Perin2011}. Not every connection is equally likely to be established, rather some pairs of neurons are more likely connected than others. In this context, the relative occurrence of bidirectionally connected pairs has been of particular interest. Using data obtained from paired whole-cell recordings in cortical slices, the amount of bidirectionally connected pairs was compared to the number of reciprocal pairs as one would expect in a random network with the same overall connection probability. Connectivity of layer 5 pyramidal neurons in the rat visual cortex \cite{Song2005} and somatosensory cortex \cite{Markram1997, Perin2011} was shown to have a significantly stronger reciprocity than expected.


The prevalence of bidirectional connectivity has since been established as an important indicator for the non-randomness of a network \cite{Lefort2009, Bourjaily2011}. However, the exact relationship between non-randomness and relative reciprocity has not been explained. Here, we model cortical circuitry as random networks in which each possible connection has a separate probability to exist. Using this model we're able to show that any non-random connectivity, expressed as higher connection probabilities in some edges and lower probabilities in others, necessarily induces a relative overrepresentation of bidirectional connections as long as connection probabilities remain symmetric within pairs. Quantitatively, we analyze reciprocity in networks with a discrete and a continuous distribution in connection probabilities to demonstrate that a relative occurrence of bidirectional connections as reported from experimental studies can be easily obtained in these models.







\section*{Results}





The emergence of non-random connectivity patterns can be modeled by
assigning each possible connection in a random graph a separate
probability to exist.
In such a model some connections are more likely to be realized than
others, allowing for the encoding of patterns within the specific
probabilities of each connection.
In the limiting case each connection either exists or is absent with
certainty, representing a blueprint for the network architecture.
To analyze the effect of non-random structures within a network,
specifically on the statistics of bidirectionally connected pairs
found in the network, we consider a random graph model of $N$ neurons
in which the probability of node $i$ to connect to node $j$ is modeled by a random variable $P_{ij}$.
For this we assume $P_{ij}$ for $i,j = 1,\dots,N$ with $i \neq j$ to be
identically distributed random variables in $[0,1]$, yielding a
probability of connection for each ordered pair of nodes in the graph.
Outside of pairs the random variables $P_{ij}$ are assumed to be independent, that is non-equal $P_{ij}$ and $P_{kl}$ are independent as long as $i \neq l$ or $j \neq k$.
Finally, we explicitly exclude self-connections in this model and assume at all
times that $i \neq j$.
%

%
Given the distributions of connection probabilities, what is
then the probability in this model for a randomly selected node to have a
projection to another randomly selected node?
As the random variables $P_{ij}$ are identically distributed, we
compute this overall connection probability $\mu$ easily as the
expected value of~$P_{ij}$,
\begin{align}
\mu = \E(P_{ij}).
\end{align}
For example, if the $P_{ij}$ have a probability density function $f$ with essential support in $[0,1]$, we can compute the connection fraction as
\begin{align}
  \mu = \int_0^1 x f(x)\,dx.
\end{align}

In this work we are interested in the probability $P_{\mathbf{bidir}}$
of a bidirectional connection to exist in a random pair of neurons.
We determine $P_{\mathbf{bidir}}$ as the expected value of the product
of $P_{ij}$ and $P_{ji}$,
\begin{align}
P_{\mathbf{bidir}} = \E(P_{ij} P_{ji}).
\end{align}
The relative occurrence $\varrho$ of such reciprocally connected pairs compares $P_{\mathbf{bidir}}$ with the occurrence of bidirectional pairs in an Erd\H{o}s-R\'{e}nyi graph, in which each unidirectional connection is equally likely to occur with probability $\mu$ \cite{Gilbert1959, Erdos1959}. The probability of a particular bidirectional connection to exist in such a random graph is simply $\mu^2$ and we obtain the relative occurrence as the quotient
\begin{align}
\varrho = \frac{P_{\mathbf{bidir}}}{\mu^2} = \frac{\E(P_{ij}P_{ji})}{{\E\left(P_{ij}\right)}^2}.\label{eq:rho}
\end{align}
Experimental studies in local cortical circuits of rodents have repeatedly reported a relative occurrence of bidirectional connections $\varrho > 1$ \cite{Markram1997, Song2005, Perin2011}. To understand in which cases such an overrepresentation occurs, we consider two cases. In the first case, assume that connection probabilities are independently determined in pairs as well, meaning that the random variables $P_{ij}$ and $P_{ji}$ are independent. Then, as $P_{ij}$ and $P_{ji}$ are identically distributed,
\begin{align}
\E(P_{ij} P_{ji}) = \E(P_{ij})\,\E(P_{ji}) = \E(P_{ij})^2, 
\end{align}
and we expect to observe no overrepresentation of reciprocal connections, $\varrho = 1$.  In the second case, assume that connection probabilities are symmetric in pairs, $P_{ij} = P_{ji}$. In this case,
\begin{align}
P_{\mathbf{bidir}} = \E(P_{ij}^2),
\end{align}
and the expected relative occurrence of reciprocal connections becomes
\begin{align}
\varrho = \frac{\E(P_{ij}^2)}{{\E\left(P_{ij}\right)}^2}.
\end{align}
We note that now any distribution of $P_{ij}$ with a nonvanishing
variance will lead to a relative occurrence that deviates from the
Erd\H{o}s-R\'{e}nyi graph, as
\begin{align}
\Var(P_{ij}) = \E(P_{ij}^2) - \E\left(P_{ij}\right)^2.
\end{align}
Moreover, since $x \mapsto x^2$ is a strictly convex function, Jensen's inequality \cite{Jensen1906, Cover2006} yields
\begin{align}
\E(P_{ij}^2) \geq \E(P_{ij})^2, \label{eq:jensen}
\end{align}
and we find that $\varrho \geq 1$ in networks with symmetric connection probabilities. Jensen's inequality further states that equality in \eqref{eq:jensen}, and thus $\varrho = 1$, holds if and only if $P_{ij}$ follows a degenerate distribution, that is if all $P_{ij}$ take the identical value $\mu$. In the other case, where the $P_{ij}$ take on more than one value with non-zero probability, we speak of a non-degenerate distribution.

As a central result of this study we thus find that any non-degenerate distribution of symmetric connection probabilities ($P_{ij} = P_{ji}$) necessarily induces an overrepresentation of bidirectional connections in the network, $\varrho > 1$. In other words, in a network in which both directions of connection are equally likely within any given pair, but where some pairs are more likely to be connected than others, the count of expected reciprocally connected pairs is strictly underestimated by the statistics of an Erd\H{o}s-R\'{e}nyi graph with same the overall connection probability $\E(P_{ij}) = \mu$.




  \subsection*{Upper bound for $\varrho$}

The overrepresentation of bidirectional connections $\varrho$ in a network is maximal when every connected pair is already a reciprocally connected pair. In terms of the model defined above, this is the case when
\begin{align}
\E(P_{ij} P_{ji}) = \E(P_{ij}).
\end{align}
The relative occurrence of reciprocal connections from \eqref{eq:rho} then becomes
\begin{align}
\varrho = \frac{1}{\E(P_{ij})} = \frac{1}{\mu} \label{eq:rho_max}
\end{align}
Thus, for local cortical circuits of L5 pyramidal neurons with a typical connection probability of $\mu = 0.1$ \cite{Thomson2002,Song2005}, the network model yields a maximal overrepresentation of $\varrho = 10$. While this theoretical maximum is unlikely to exist in actual cortical networks, the precise degree of overrepresentation will depend on the specific distribution of connection probabilities in the network. In the following, we study two generic examples.

  \subsection*{Two-point distribution}

\begin{figure}[h!]
\centering
\includegraphics[width=\textwidth]{%
    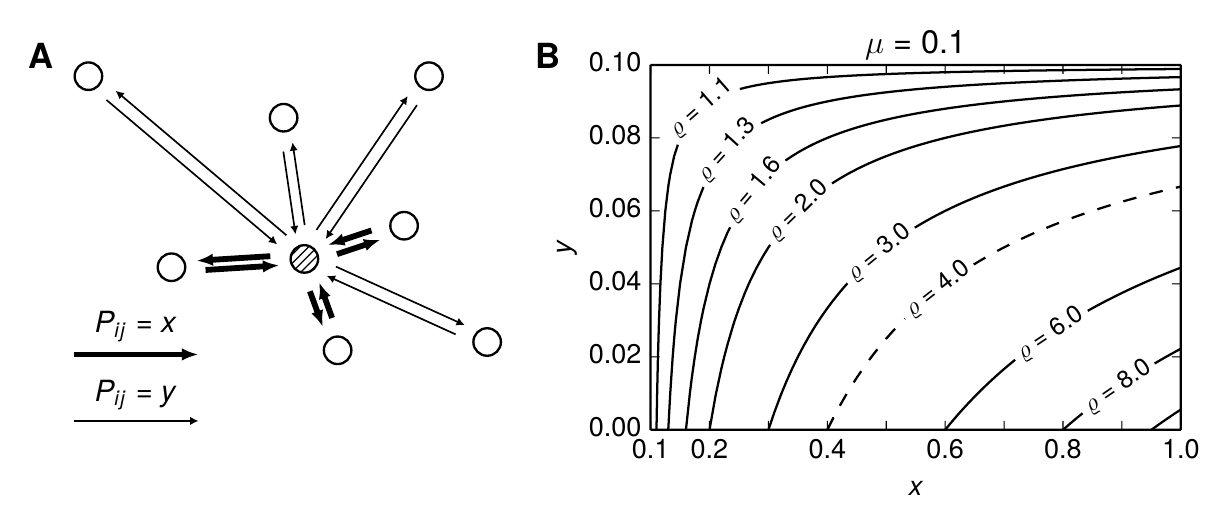}
\caption{Relative overrepresentation $\varrho$ of bidirectional
  connections in networks with a fraction of pairs connected with a
  high probability $x$ and the rest of the pairs connected with a low
  probability $y$. \textbf{A}~ Diagram illustrating the targets
  connecting with a high chance $x$ (thick arrows) and targets
  connecting with a low probability $y$ (thin arrows) for a single
  source node (hatched). \textbf{B}~ Different pairings of $x$ and $y$
  can induce a high relative overrepresentation $\varrho$ in a network
  with two-point distributed connection probabilities, $P_{ij} \sim
  \mathcal{T}(\frac{\mu-y}{x-y},x,y)$, and a fixed overall connection
  probability $\mu = 0.1$. The dashed line marks an overrepresentation
  of bidirectional connections of $\varrho=4$ as observed for layer 5
  pyramidal neurons in the rat visual cortex \cite{Song2005}.}
\label{fig:tp}
\end{figure}

The simplest non-degenerate distribution of connection probabilities
is a distribution that takes two values $x$, $y$ with probability $p$
and $1-p$, respectively, as illustrated in Figure~\ref{fig:tp}A. %
This distribution may be seen as a crude approximation to the
connection probabilities recently observed in visual cortex as a
function of the neurons' absolute difference in orientation preference, where a \enquote{high} connection probability was reported for a difference between 0\textdegree--45\textdegree\ and a \enquote{low} probability was seen for cells with a difference of 45\textdegree--90\textdegree\ in orientation tuning \cite{Lee2016a}.
Formally, let $x,y \in [0,1]$ with $x > y$ and $0 < p
< 1$. A random variable $X$ follows the two-point distribution 
$\mathcal{T}(p,x,y)$ if $P(X=x)=p$ and $P(X=y) = 1-p$.
In our network model let then the $P_{ij}$ be $\mathcal{T}(p,x,y)$
distributed. The overall connection probability $\mu$ is
\begin{align}
\mu = \E(P_{ij}) = px + (1-p)y. \label{eq:bd1}
\end{align}
Assume again that $P_{ij} = P_{ji}$. The relative occurrence of
bidirectional connections is given by
\begin{align}
  \varrho = \frac{\E(P_{ij}^2)}{\mu^2} = \frac{p x^2 + (1-p) y^2}{\mu^2} \label{eq:bd2}.
\end{align}
Solving \eqref{eq:bd1} for $p$ as
\begin{align}
p = \frac{\mu - y}{x-y}
\end{align}
and inserting into equation~\eqref{eq:bd2} yields an expression for
the relative overrepresentation depending on $x$, $y$ and $\mu$ (see
Supplementary Information SI1),
\begin{align}
\varrho = \frac{x+y}{\mu} - \frac{xy}{\mu^2}.
\end{align}

%
Here we fix $\mu = 0.1$ in accordance with the overall connection
probability found in local circuits of pyramidal cells in the rat visual
cortex \cite{Song2005} and obtain the relative occurrence dependent
on the two connection probability values $x$ and $y$.
Given $x \geq \mu$ it follows that $y \leq \mu$ (see Supplementary
Information SI2) and the possible values for $x$ and $y$ are $0.1 \leq x
\leq 1$ and $0 \leq y \leq 0.1$.
Figure~\ref{fig:tp}B shows contours of $\varrho$ for the $(x,y)$
pairings illustrating how different values for the relative
overrepresentation of reciprocal connections can be induced by
two-point distributed connection probabilities.
We find that in such networks higher values of $\varrho$ are easily
obtained with reasonable network configurations. %
For example, a relative overrepresentation of $\varrho=4$ could be
achieved by a two-point distribution of connection probabilities where
one group of neuron pairs is highly connected with probability
$x=0.7$, while the other group of neuron pairs is sparsely connected
with probability $y=0.05$.
Collectively, the highly connected pairs then make up less than $8\%$
of all neuron pairs, showing that it is sufficient to have a small
subgroup of highly connected neuron pairs to induce a high
overrepresentation of bidirectionally connected pairs in the
network. For more densely connected networks, $\mu>0.1$, the effect that
two distinct connection probabilities have on the overrepresentation
of reciprocal connections is reduced (c.f. Figure~S1), as one would
intuitively expect from the dependence of the maximal
overrepresentation on $\mu$ in \eqref{eq:rho_max}.


  \subsection*{Gamma distribution}

Next, we analyze the relative overrepresentation of bidirectional
connections in a network with continuously distributed connection
probabilities. The gamma distribution $\Gamma(\alpha, \beta)$ with
probability density function
\begin{align}
    f_{\alpha,\beta}(x) = \begin{cases} 
\frac{1}{\beta^{\alpha}\Gamma(\alpha)}\, x^{\alpha-1}\,e^{-x/\beta} & x \geq 0 \\
0 & \text{otherwise},
\end{cases}
\end{align}
allows the variation of the variance $\Var(X)= \alpha \beta^2 $ of a
gamma distributed random variable $X \sim \Gamma(\alpha, \beta)$, while
keeping its mean $\E(X) = \alpha \beta $ constant \cite{Hogg1978}.
The exponential distribution emerges as a special case of the gamma
distribution ($\alpha =1$).

\begin{figure}[h!]
\centering
\includegraphics[width=\textwidth]{%
  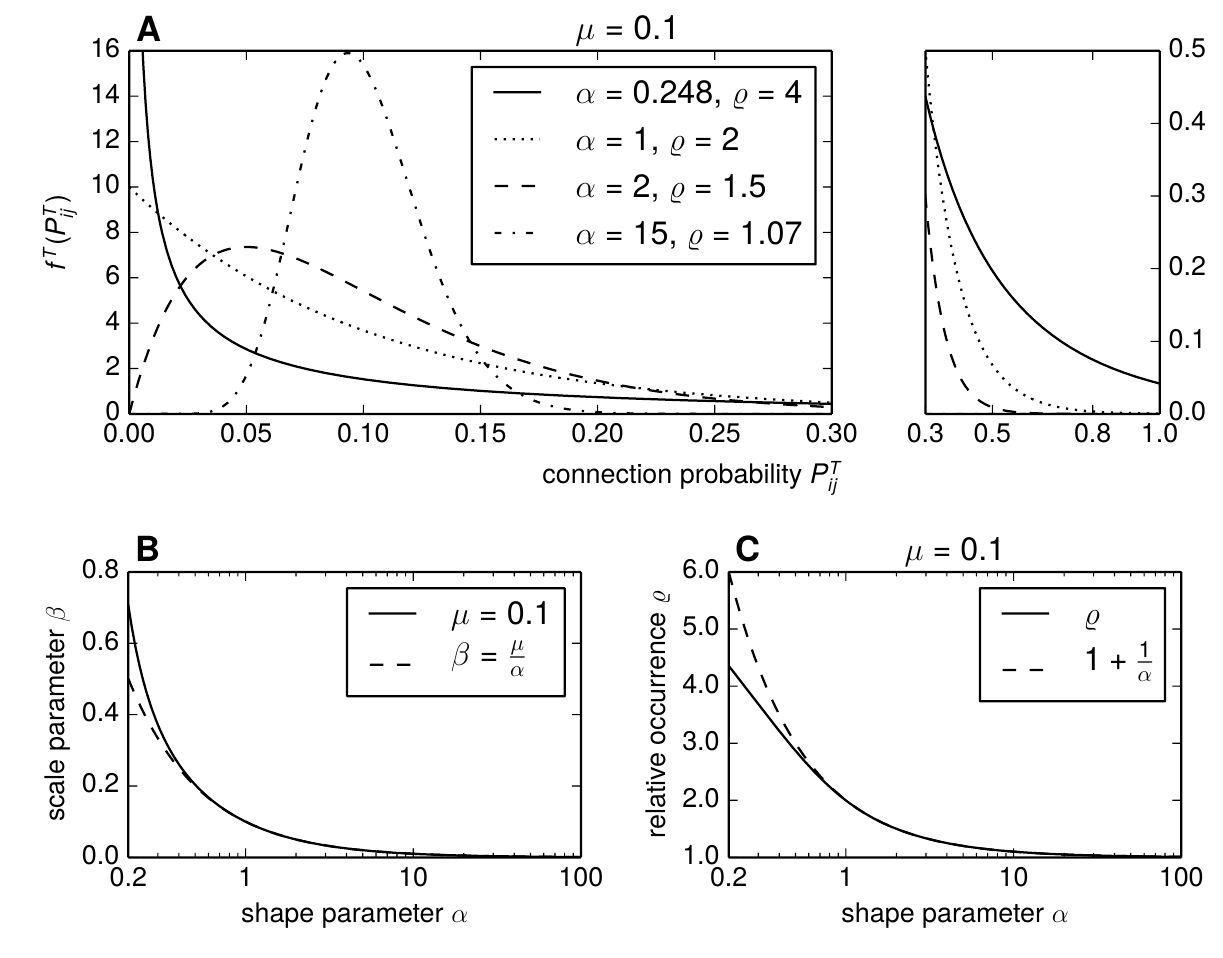}
\caption{Relative occurrence of bidirectional connections $\varrho$ in
  networks with gamma distributed connection probabilities. \textbf{A}
  Probability density functions of the truncated gamma distribution
  $\Gamma^T(\alpha,\beta)$ for different shape parameters $\alpha$ and
  the induced relative overrepresentation $\varrho$ in a network with
  such distributed connection probabilities $P_{ij}$. For a given
  $\alpha$, the scale parameter $\beta$ was chosen such that $\mu =
  0.1$. Plot to the right continues the density functions on a
  different scale. \textbf{B} Contour of $\alpha$, $\beta$ pairings
  that yield an overall connection probability of $\mu = 0.1$. The
  dashed line shows the approximation $\beta=\frac{\mu}{\alpha}$,
  where $\mu=0.1$. \textbf{C} Relative occurrence $\varrho$ as a
  function of $\alpha$ for fixed $\mu = 0.1$. For $\alpha \geq 1$ this
  relationship is well approximated by $\varrho \approx 1 +
  \frac{1}{\alpha}$.}
\label{fig:gd}
\end{figure}

To ensure that the randomly drawn connection probabilities lie within the
interval $[0,1]$, we here consider a modification to the
traditional gamma distribution in the form of a truncated version. Let
$\alpha, \beta > 0$. A random variable $X$ follows the truncated gamma
distribution $\Gamma^T(\alpha, \beta)$ if it has the probability
density function
\begin{align}
  f_{\alpha,\beta}^T(x) = \begin{cases} K_{\alpha, \beta}\,
\frac{1}{\beta^{\alpha}\Gamma(\alpha)}\, x^{\alpha-1}\,e^{-x/\beta} & 0 \leq x \leq 1 \\
0 & \text{otherwise}.
\end{cases} \label{eq:fTab}
\end{align}
The factor $K_{\alpha,\beta}$ is the inverse of the cumulative
probability that $x \leq 1$ of the untruncated gamma distribution,
\begin{align}
  K_{\alpha,\beta} = \left(\int_0^{1} f_{\alpha,\beta}(x) \, dx \right)^{-1},
\end{align}
and is needed to ensure that
\begin{align}
  \int f_{\alpha,\beta}^T(x) \,dx = 1 \label{eq:gd1}.
\end{align}
Consider then the above network model in which the connection
probabilities $P_{ij}^T$ are $\Gamma^T(\alpha, \beta)$ distributed and $P_{ij}^T = P_{ji}^T$. We
compute the relative overrepresentation $\varrho$ numerically from
\begin{align}
  \mu = \E\left(P_{ij}^T\right) &= \int_0^1 x f_{\alpha,\beta}^T(x)\, dx,\\
        \E\left({P_{ij}^T}^2\right) &= \int_0^1 x^2 f_{\alpha,\beta}^T(x)\, dx.
\end{align}
Pairings of the shape parameter $\alpha$ and the scale parameter
$\beta$ were chosen such that the overall connection probability
reflects connectivity statistics in local cortical networks, $\mu =
0.1$ \cite{Song2005, Thomson2002}. Probability density functions and
resulting relative overrepresentation of reciprocal connections
$\varrho$ for four representative $\alpha,\beta$ pairs are shown in
Figure~\ref{fig:gd}A. Here, $\beta$ was determined to yield $\mu =
0.1$ for the given $\alpha$, following the relationship shown in
Figure~\ref{fig:gd}B (solid curve).


In the sparse networks we modeled, the tail of the gamma distribution
is near zero at $P_{ij}^T=1$ (see Figure 2A). Thus $K_{\alpha, \beta}
\approx 1$ and the truncated gamma distribution can be well
approximated by the untruncated version. Assuming connection
probabilities to be standard gamma distributed, $P_{ij} \sim
\Gamma(\alpha,\beta)$, we have
\begin{align}
  \E(P_{ij}^2) = \Var(P_{ij}) + \E(P_{ij})^2 =  \alpha \beta^2 + \alpha^2 \beta^2, 
\end{align}
and thus
\begin{align}
  \varrho = \frac{\E\left({P_{ij}^T}^2\right)}{\E\left(P_{ij}^T\right)^2} \approx \frac{\E\left({P_{ij}}^2\right)}{\E\left(P_{ij}\right)^2} =  \frac{\alpha^2 \beta^2}{\alpha^2 \beta^2} + \frac{\alpha \beta^2}{\alpha^2 \beta^2} =
 1 + \frac{1}{\alpha}  =: \tilde{\varrho}.
\end{align}
The approximation $\varrho \approx \tilde{\varrho} = 1 + \frac{1}{\alpha}$ works well for
$\alpha \geq 1$ as shown in Figure~\ref{fig:gd}C.





To induce a high overrepresentation of reciprocal pairs in the
network, the gamma distribution of connection probabilities takes a
highly skewed shape. In order to obtain $\varrho = 4$, only $57\%$ of
pairs are expected to have a higher connection probability than $0.01$
($\alpha=0.248$, $\beta = 0.487$). Such a situation in which a large
part of all neuron pairs have a small connection probability while
some few pairs have a high chance to be connected is likely if, e.g.,
the connection probability strongly depends on the spatial separation
of the neuron, as it was found in layer 5 excitatory circuits of the
rat somatosensory cortex \cite{Perin2011}. Then only nearby neurons
are likely to be connected, while the larger part of more distant
neurons has a low probability of connection.

  \subsection*{Symmetry of connection probabilities in neural circuits}


In neural circuits, connection probabilities that are equal within pairs but differ across the network are plausible from both an anatomical and a functional perspective. From the anatomical point of view, the distance dependency of connection probabilities mentioned above is a characteristic of cortical circuits that necessarily leads to symmetric probabilities: the distance from the first neuron's soma to the second neuron's soma is the same as the distance from the second to the first, resulting in equal probabilities within a pair of neurons when inter-neuron distance determines connection probabilities. Regarding the functional perspective, connection probabilities may also depend on functional properties of the cells in the network. For example, the probability of connection of orientation tuned cells in the mouse primary visual cortex depends on their absolute difference in orientation tuning \cite{Lee2016a, Ko2011}. Since the absolute difference in orientation tuning will be the same in both directions, connection probabilities can be expected to be equal within a pair of orientation tuned cells.

\begin{figure}[h!]
\centering
\includegraphics[width=\textwidth]{%
  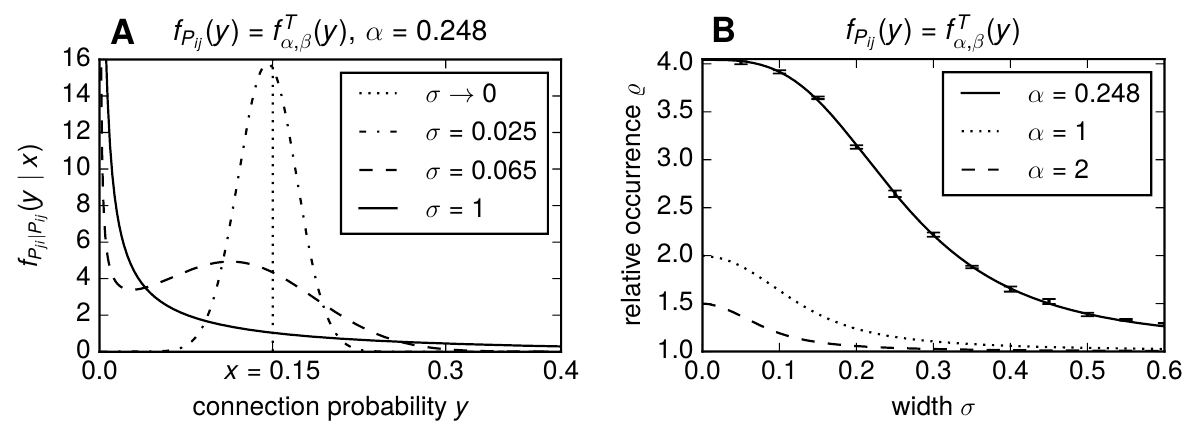}
\caption{Relative overrepresentation of bidirectional connections $\varrho$ is sustained when connection probabilities are only approximately symmetric in pairs. \textbf{A} Illustration how the conditional density function $f_{P_{ji} | P_{ij}} (y\,\vert\, x)$ of \eqref{eq:fpijpji} transitions from equality of the random variables $P_{ij}$ and $P_{ji}$ to independence with increasing $\sigma$. We use $f_{P_{ij}} (y) = f^T_{\alpha,\beta}(y)$, with $\alpha=0.248$ and $\beta$ such that $\operatorname{E}(P_{ij})=0.1$. For the illustration $P_{ij}$ was fixed as $x=0.15$. Already for $\sigma=1$ the conditional density function becomes visually indistinguishable from  $f^T_{\alpha,\beta}(y)$. \textbf{B}~Relative occurrence of reciprocally connected pairs $\varrho$ as a function of $\sigma$. The curves for $\alpha=1$ and $\alpha=2$ show numerical solutions of \eqref{eq:rho_sigma} with $f_{P_{ij}}(y) = f_{\alpha,\beta}^T(y)$, where $\beta$ was chosen such that $\operatorname{E}(P_{ij}) = 0.1$. Relative reciprocal pair counts from generated networks following the model matched these theoretical curves (data not shown). For $\alpha = 0.248$ random variables with the respective probability density functions were sampled and the average $\varrho$ was computed via \eqref{eq:rho_sigma} using the sample means. Error bars show SEM, the curve for $\alpha = 0.248$ (solid line)  was fitted to the data points and is purely for illustrative purposes.} 
\label{fig:sym_fail}
\end{figure}

However, even when connection probabilities within pairs do not match exactly, an overrepresentation of reciprocal connections is still likely to be observed when connection probabilities follow a non-degenerate distribution. To see this, consider that connection probabilities $P_{ij}$ are distributed according to some probability density function $f_{P_{ij}}(x)$. As before we assume that the $P_{ij}$ are independent outside of pairs. In the following we also assume that $i > j$ without loss of generality. The expected probability of a reciprocal connection within a pair can then be expressed as
\begin{align}
  \E(P_{ij}P_{ji}) = \int_0^1 \int_0^1 xy\, f_{P_{ij},P_{ji}}(x,y) \, dx\, dy, \label{eq:dbint}
\end{align}
where $f_{P_{ij},P_{ji}}(x,y)$ is the joint probability density function of $P_{ij}$ and $P_{ji}$, 
\begin{align}
  f_{P_{ij},P_{ji}}(x,y) =  f_{P_{ji} | P_{ij}}(y \mid x) f_{P_{ij}}(x). \label{eq:cdf_def}
\end{align}
In the case that $P_{ij}$ and $P_{ji}$ are independent we have $f_{P_{ji} | P_{ij}}(y \mid x) = f_{P_{ji}}(y)$ and in the case of $P_{ij}=P_{ji}$ it is $f_{P_{ji} | P_{ij}}(y \mid x) = \delta(y-x)$. Here we propose a model for the conditional density function that transitions between the two extreme cases by multiplying $f_{P_{ji}}(y)$ with the density function of a normal distribution centered around $x$,
\begin{align}
  f_{P_{ji} | P_{ij}} (y \mid x) = \frac{1}{N_{\sigma}(x)} f_{P_{ji}}(y)\, \frac{1}{\sigma \sqrt{2 \pi}} \,e^{\frac{(y-x)^2}{2 \sigma^2}} \label{eq:fpijpji},
\end{align}
where the additional factor $N_{\sigma}(x)^{-1}$  makes sure that $f_{P_{ji}|P_{ij}} (y \mid x)$ integrates to one,
\begin{align}
  N_{\sigma}(x) = \int_0^1 f_{P_{ji}}(z)\, \frac{1}{\sigma \sqrt{2 \pi}}\, e^{\frac{(z-x)^2}{2 \sigma^2}} \,dz.
\end{align}
Indeed, as the standard deviation $\sigma$ of the modulating normal distribution increases $f_{P_{ji}|P_{ij}} (y \mid x)$ approaches $f_{P_{ji}}(y)$ and in the limit $\sigma \to 0$ we have
\begin{align}
  \lim_{\sigma \to 0}   f_{P_{ji}|P_{ij}} (y \mid x) = \delta(y-x).
\end{align}
In Figure~\ref{fig:sym_fail}A, conditional density functions for various $\sigma$ are shown for the truncated gamma distribution. For low values of $\sigma$ the conditional density function resembles a narrow Gaussian around $x$, reflecting approximately symmetric connection probabilities. For $\sigma > 1$ on the other hand $f_{P_{ji} | P_{ij}}(y \mid x)$ becomes virtually indistinguishable from $f^T_{\alpha, \beta}(y)$, reflecting independence of $P_{ji}$ from $P_{ij}$.

Finally we employ the model to examine how the relative overrepresentation of bidirectional connections $\varrho$ changes with the degree of symmetry in the connection probabilities within a pair of neurons. For this $\varrho$ is computed as a function of $\sigma$ as for a given distribution of $P_{ij}$ as
\begin{align}
  \varrho = \frac{\E(P_{ij} P_{ji})}{\mu^2}, \label{eq:rho_sigma}
\end{align}
where the numerator is given by \eqref{eq:dbint} with 
\eqref{eq:fpijpji} 
%
%
and the overall connection probability $\mu$ is calculated as
\begin{align}
 \mu = \frac{1}{2} \int_0^1 x f_{P_{ij}}(x)\,dx + \frac{1}{2} \int_0^1 f_{P_{ij}}(x) \int_0^1 y \,f_{P_{ji}\vert P_{ij}}(y \mid x) \,dy \, dx.
\end{align}

Figure~\ref{fig:sym_fail}B shows the change of $\varrho$ with $\sigma$ for connection probabilities $P_{ij}$ following a truncated gamma distribution $\Gamma^T(\alpha, \beta)$. For the three parameter sets chosen we see that a strong overrepresentation of bidirectional connections is sustained when connection probabilities are only approximately symmetric in pairs. Furthermore, as long as $P_{ji}$ is at least somewhat biased to take similar values to $P_{ij}$ an overrepresentation of $\varrho > 1$ can be observed, implying that effects such as distance-dependency or the dependence on the absolute difference in orientation tuning of connection probabilities will tend to increase the relative occurrence of bidirectional connections, even when other effects are also influencing the neurons' connection probabilities.

%

\section*{Discussion}


Experimental evidence suggests that any pair of excitatory cells within a cortical column has contact points between axon and dendrite close enough to support a synaptic connection between the cells \cite{Stepanyants2004,Kalisman2005}.
Despite this potential \enquote{all-to-all} connectivity, only a small fraction of the contacts are realized as functional synapses.
Uncovering the underlying principles of which contact points get utilized for synaptic transmission is crucial for our understanding of the structure and function of the local cortical circuits in the mammalian brain.

The emerging local networks in the rat visual and somatosensory cortex have been shown to feature non-random structure \cite{Song2005, Perin2011} and much attention was given to bidirectionally connected neuron pairs that are occurring more often than expected from random connectivity \cite{Bourjaily2011, Clopath2010, Miner2016}.
In this study we have shown a condition under which non-random network structure and the occurrence of reciprocally connected pairs are inherently linked; a relative overrepresentation of bidirectional connections arises necessarily in networks with a non-degenerate distribution of symmetric connection probabilities.
Absence of an overabundance of reciprocal pairs on the other hand, as for example found in the intra-layer connectivity of the mouse C2 barrel column \cite{Lefort2009}, points towards either a truly random network or an asymmetry in the connection probabilities.

Quantitatively, a network in which connection probabilities take on one of two values is easily able to account for even the highest values of overrepresentation reported.
A network with such a two-point distribution of connection probabilities might occur naturally, where the probability of connection depends on whether a given pair of neurons shares a certain feature, for example has a similar orientation preference or not \cite{Lee2016a}.

A continuous distribution in connection probabilities on the other hand might occur when pair connectivity depends on a continuous parameter such as the inter-neuron distance or the neurons' age.
We showed that networks in which connection probabilities follow a gamma distribution can as well have a high relative occurrence of reciprocally connected pairs, however in this case a larger fraction of pairs remain unconnected with a very high probability.

It is likely that a combination of such effects determines the connection probabilities in local cortical networks.
Importantly, we showed that as long as this probability is symmetric for pairs, any such effect that creates a non-degenerate distribution of probabilities will cause an increase of the reciprocity in the network.

%

Our results confirm the intuitive notion that reciprocity is favored in symmetric networks, whereas asymmetric probabilities of connection inhibit the occurrence of bidirectionally connected pairs. Network models with symmetric connectivity such as Hopfield nets generally excel at memory storage and retrieval through fixed point attractor dynamics \cite{Hopfield1982}, while asymmetric network models such as synfire chains are suitable for reliable signal transmission \cite{Abeles1982, Diesmann1999}. This suggests the intriguing possibility that one may be able to infer the nature of the computations in a neural circuit based on certain statistics of its connectivity such as the abundance of bidirectionally connected pairs.

In conclusion, the present study puts the overrepresentation of bidirectional connections found in local cortical circuits in a new light.
If connection probabilities are symmetric in pairs, the overrepresentation emerges as a symptom of any form of non-random connectivity.
It is thus crucial for both future experimental and modeling studies to develop a more refined view of non-random network connectivity that goes beyond simple pair statistics. Focusing on higher order connectivity patterns and taking into account the actual synaptic efficacies seem promising avenues for future research into the non-random wiring of brain circuits.

\newpage
\section*{Supplementary Material}
The supplementary information document for references SI1 and SI2 and Figure S1 is available online at \textsc{doi}:~\texttt{\href{https://dx.doi.org/10.6084/m9.figshare.3501860}{\nolinkurl{10.6084/m9.figshare.3501860}}}. Python code for the numerical computations is available as a GitHub repository and was archived including the generated data at \textsc{doi}:~\texttt{\href{https://doi.org/10.5281/zenodo.200368}{\nolinkurl{https://doi.org/10.5281/zenodo.200368}}}. A website documenting the code is found at \texttt{\href{https://non-random-connectivity-comes-in-pairs.github.io/}{\nolinkurl{https://non-random-connectivity-comes-in-pairs.github.io/}}}.

\section*{Acknowledgements}
The authors would like to thank the anonymous reviewers for their helpful and constructive comments on earlier versions of this article. JT is supported by the Quandt foundation.




\printbibliography

\end{document}